\documentstyle[11pt,twoside,fancyhea,cmp]{article}

\title{\Large\bf
RELAXATION DYNAMICS OF DISORDERED ISING MODEL.
TWO-SITE CLUSTER APPROXIMATION
   }
\author{
   {\sc R.~R.~Levitskii, S.~I.~Sorokov, R.~O.~Sokolovskii}
   \\[1.5ex]
   \it Institute for Condensed Matter Physics
   \\
   \it of the Ukrainian National Academy of Sciences
   \\
   \it 1~Svientsitski St., UA--290011 Lviv, Ukraine
}
\date{Received September 11, 1995}

\newcommand{\be}{\begin{equation}}
\newcommand{\ee}{\end{equation}}
\newcommand{\braced}[1]{\left\{#1\right\}}
\newcommand{\ptsed}[1]{\left(#1\right)}
\newcommand{\ang}[1]{\langle #1\rangle}
\newcommand{\ab}{_{\alpha\beta}}
\newcommand{\ia}{_{i\alpha}}
\newcommand{\jb}{_{j\beta}}
\newcommand{\iajb}{_{i\alpha,j\beta}}
\newcommand{\bee}{\begin{eqnarray}}
\newcommand{\eee}{\end{eqnarray}}
\newcommand{\fr}{ \, _r\bar{\varphi}}
\newcommand{\kappat}{\bar{\kappa}}
\newcommand{\kt}[1]{\, _{#1}\kappat}
\newcommand{\hafitz}[1]{ \hat{\phi}^{(20)}_{#1}  }
\newcommand{\hafit}[1]{ \hat{\phi}^{(2)}_{#1} }
\newcommand{\hafioo}[1]{ \hat{\phi}^{(11)}_{#1} }

\newcommand{\avS}[1]{\langle #1\rangle_{H,t}}
\newcommand{\avX}[1]{\langle #1\rangle_x}
\newcommand{\Piq}[1]{ \pi_{\vec{#1}} }
\newcommand{\dif}[2]{\frac{d #1}{d #2}}
\newcommand{\fdif}[2]{\frac{\delta #1}{\delta #2}}
\newcommand{\sect}[1]{\section{#1}\setcounter{equation}{0}}
\renewcommand{\theequation}{\arabic{section}.\arabic{equation}}

\begin{document}\maketitle
\begin{abstract}
{\small
Spin relaxation in a site-disordered Ising model within
master equation approach is studied. The $\vec{q},\omega$-dependent
susceptibility of the model is calculated and investigated.
Effects described by the two-site cluster approximation and
lost by the mean field approximation are discussed. Comparison of obtained
results with dielectric measurements in $Cs(H_{1-x}D_x)_2PO_4$ is presented.
}\end{abstract}

\sect{Introduction}

       Being  the  model  of  critical  phenomena  in many different objects
an Ising model  compels  an  attention  during  many  years. This paper
investigate the kinetics of the model  (within master equation \cite{Glauber}
approach) having several goals. We  consider the disordered  (site chaos)
Ising  model in order to describe  the dynamic  properties of  partially
deuterrated  ferroelectrics with  hydrogen  bonds.  Besides  the  theory
seems  applicable  to many other systems:  alloys  of  magnets,  diluted
magnetic or ferroelectric systems and others. Within two-site  cluster
approximation (TCA)  we calculate the  dynamic stucture factor to supplement
our previous study \cite{cmp1}  of correlation functions of the model.

\sect{The Hamiltonian. The approach}

We consider the spin system with site chaos on the Bravais lattice which is
described by the Ising Hamiltonian
\be
 H = -\sum_i\kappa_i S_i - \frac{1}{2}\sum_{ij} K_{ij}S_i S_j \;.
\label{Hamilt}
\ee
$H$ describes the set of Ising spins with the pair exchange interaction
$K_{ij}$ in the site-dependent field $\kappa_i$. The set of variables
$\braced{S_i}$ ($S_i=\pm 1$; $i=1\cdots N$, $N$ is a number of the sites)
represents a state of spin subsystem, and the set $\braced{X\ia}$ describes
the sort configuration ($X\ia=1$, if the site $i$ is occupied by spin of
the sort $\alpha$, otherwise $X\ia=0$; $\alpha=1\cdots\Omega$, $\Omega$ is a number
of the sorts). The quantities $\kappa_i$, $K_{ij}$ depend on sort
configuration
\be
\kappa_i = \sum_\alpha \kappa\ia X\ia;\; K_{ij} = \sum\ab K\iajb X\ia X\jb
\ee
so the true parameters of the Hamiltonian (\ref{Hamilt}) are $\kappa\ia$,
$K\iajb$. Here we suppose the pair interactions to be short-range
\be
K\iajb = K\ab\pi_{ij};\;K\ab>0;\;
 \pi_{ij} = \left\{
\begin{array}{ll}
        1, & $if $j\in\pi_i\\
        0, & $otherwise$           
\end{array}
             \right.
\ee
where $ \pi_i $ denotes the set of the nearest neighbours of the site $i$ (the
first coordination sphere).

From the very beginning we shall consider the quenched system (frozen sort
configuration) and investigate its dynamics on the basis of Glauber's master
equation \cite{Glauber} for the distribution function $\rho(\braced{S},t)$:
\bee
  \dif{ }{t} \rho(\{S\},t) &=&
 -\sum_i W_i(\cdots S_i \cdots ) \rho(\cdots S_i\cdots ,t) \nonumber\\
&& +\sum_i W_i(\cdots -S_i\cdots ) \rho (\cdots -S_i\cdots ,t) 
\label{2.4}
\eee
Following ref. \cite{Glauber} the probability per unit time that the $i$th
spin flips from the value $S_i$ to $-S_i$ is assumed to be
\be
W_i(\cdots S_i\cdots ) = \frac{1}{2\tau_i}(1 - S_i \tanh\beta\varepsilon_i)
\label{2.5}
\ee
where
\be
     \varepsilon_i= \kappa_i+\sum_jK_{ij}S_j
\label{2.6}
\ee
is a local effective field acting on the $i$th spin. In this approach the
dissipative properties of the spin subsystem is caused by its interaction
with another (e.g. phonon) subsystems which play the role of a heat bath. The
quantity $\tau_i$ has the dimension of time and defines the time scale of
relaxational phenomena.  We assume the quantity $\tau_i$ to be locally
sort-dependent:
\be
\tau_i=\sum_\alpha \tau_\alpha^0 X\ia
\ee
Relations (\ref{2.4}) and (\ref{2.5}), (\ref{2.6}) lead to the following
kinetic equations for the average value of spin:
\be
    D_{i,t}\avS{S_i}= \avS{\tanh\beta\varepsilon_i}
         \label{2.9}
\ee
where
\be
  \avS{(\cdots)} = Sp_s\rho(\braced{S},t) (\cdots) ;\;
   D_{i,t} = 1 + \tau_i \dif{ }{ t}    
\ee
The system is disordered, therefore in order to calculate experimentally
measured quantities one have to perform an averaging over the sort
configurations. For example, the polarization of a partially deuterated
H-bonded ferroelectric with the Hamiltonian (\ref{Hamilt}) is defined as
\be
P(t)=\frac{1}{V}\sum_i\sum_\alpha \mu_\alpha \avX{\avS{S\ia}};\;S\ia=S_iX\ia
\ee
where $V$ is a volume of the system, $\mu_\alpha$ the dipole electric moment
associated  with the quasispin of sort $\alpha$ (proton or deuteron),
$\avX{\cdots}$ denotes the averaging over sort configurations with some
density matrix $\rho(\braced{X\ia})$ that depends on conditions of system's 
freezing. The two-site cluster approximation, which is used below, is 
sensitive only to the following moments of the distribution 
$\rho(\braced{X\ia})$
\be
\avX{X\ia}=c_\alpha,\;\avX{X\ia X\jb}=w\ab\;(j\in\pi_i)
\ee
where $c_\alpha$ has the meaning of the sort $\alpha$ spins' concentration.

\sect{Formulation of two-site cluster approximation}

The equation (\ref{2.9}) is quite intractable. One can expand the
$\avS{\tanh\beta\varepsilon_i}$ in the right-hand side of (\ref{2.9}) and find that this
equation couples correlation functions (CFs) of the type
$\avS{S_{i_1}\cdots S_{i_n}}$ where the site indices belong to the first
coordination sphere of the site $i$ ($i_1\cdots i_n \in \pi_i$). The
equation for $\avS{S_{i_1}\cdots S_{i_n}}$ derived from (\ref{2.4}) and (\ref{2.5})
involves higher CFs, so the exact treatment of the Glauber's equations has succeded
only in the case of one-dimensional system in zero field (ideal (one-sort)
chain~\cite{Glauber,SandK,Hilh}, chain with single impurity~\cite{SingleImp}). In other
cases interpolative approximations are used which express higher CFs via
lower, e.g. $\avS{\prod_j S_j}\approx\prod_j\avS{S_j}$.

        We suggest here the closure to the equation (\ref{2.9}) in the
spirit of a cluster approach. In a one-site approximation we replace
contributions of all spins with effective fields 
\be
 \varepsilon_i \rightarrow \varepsilon_i^{[1]} = \kappat_{i,t}+
 \sum_{r\in\pi_i}\fr_{i,t} \equiv \kappat_{i,t} 
\ee
where $\fr\ia$ has the meaning of effective field acting on the spin $i$ of
sort $\alpha$ from the nearest neighbour on the site $r$. In a two-site
approximation a contribution of one of the nearest neighbours is taken into
account explicitly
\be
 \varepsilon_i \rightarrow \varepsilon_i^{[2]} =
 \kappat_{i,t} + \sum_{\begin{array}{c}r\in\pi_i\\r \neq 2\end{array}}
 \fr_{i,t}+K_{ij} S_j \equiv \kt{j}_{i,t} + K_{ij} S_j,  \; (j\in\pi_i)\;.
\ee
In the case of ideal (one-sort) system no more assumptions are needed to
obtain the closed set of equations for the average value of spin
$\avS{S_i}$
\bee
 &   D_{i,t}\avS{S_i} = \avS{\tanh\beta\varepsilon_{i,t}^{[1]}} &
\nonumber\\
 &   D_{i,t}\avS{S_i} = \avS{\tanh\beta\varepsilon_{i,t}^{[2]}} \;. &
\label{3.3}
\eee
For the many-sort system it is necessary to point out a dependence of the
fields $\fr_{i,t}$ on the sort configuration. We shall believe the fields
$\fr_{it}$ to be different in the one-site approximations and the two-site
one until averaging over sort configurations is carried out. For the sort
averaging procedure both "one-site" and "two-site" fields will be assumed
to be locally sort-dependent
\be
 \fr_{i,t}= \sum_\alpha X\ia \fr_{i\alpha,t}
\ee
where $\fr_{i\alpha,t}$ is a c-number (it does not contain spin or sort
operators).

After simple transformation the equation (\ref{3.3}) can be written
in the form
\bee
 &  & D_{i,t}\avS{S_i} = \tanh\beta\kappat_{i,t} \nonumber\\
 &  & D_{i,t}\avS{S_i}=L_{ij,t}+P_{ij,t}\avS{S_j} \\
 &  & D_{j,t}\avS{S_j}=L_{ji,t}+P_{ji,t}\avS{S_i}; \qquad
       j\in\pi_i)\nonumber
\label{3.5}
\eee
where
\bee
 &
L_{ij,t}=\frac{1}{2}\left[\tanh\beta(\kt{j}_{it}+K_{ij})+
\tanh\beta(\kt{j}_{it}-K_{ij})\right]
&  \nonumber\\
&
P_{ij,t}=\frac{1}{2}\left[\tanh\beta(\kt{j}_{it}+K_{ij})-
\tanh\beta(\kt{j}_{it}-K_{ij})\right]
&
\eee
The expressions (\ref{3.5}) consist a system of nonlinear differential equations.
To linearize this expressions we consider only linear responce of the system
on a small time-dependent changing of the external field. For this purpose
we write the dynamic quantities in the following form:
\bee
&& \kappa_{i,t}=\kappa_i+\delta\kappa_{i,t};\;
\kappat_{i,t}=\kappat_i+\delta\kappat_{i,t}; \nonumber\\
&& \kt{j}_{i,t}=\kt{j}_i+\delta\kt{j}_{i,t};\;
\avS{S_i}=m^{(1)}_i+\delta m^{(1)}_{i,t}; \nonumber\\
&& L_{ij,t}=L_{ij}+L_{ij}'\;\delta\kt{j}_{i,t};\;
P_{ij,t}=P_{ij}+P_{ij}'\;\delta\kt{j}_{i,t}; \nonumber\\
&&\tanh\beta\kappat_{i,t}=
\tanh\kappat_i+(1-\tanh^2\kappat_i)\;\delta\kappat_{i,t}
\label{3.6}
\eee
Here $\kappa_i$, $\kappat_i$, $\kt{j}_i$, $m^{(1)}_i$, $L_{ij}$, $P_{ij}$ are the
static (independent of time) components of corresponding quantities and
$\delta\kappat_{i,t}$, $\delta\kt{j}_{i,t}$, $\delta m^{(1)}_{i,t}$,
$L_{ij}'\;\delta\kt{j}_{i,t}$, $P_{ij}'\;\delta\kt{j}_{i,t}$,
represent linear in $\delta\kappa_{i,t}$ deviations of these quantities from
their static values.
\bee
&&L_{ij}=\frac{1}{2}\left(\tanh\beta(\kt{j}_i+K_{ij})+
\tanh\beta(\kt{j}_i-K_{ij})\right) \nonumber\\
&&P_{ij}=\frac{1}{2}\left(\tanh\beta(\kt{j}_i+K_{ij})-
\tanh\beta(\kt{j}_i-K_{ij})\right)
\nonumber\\
&&L_{ij}'=\frac{1}{2}\left(\cosh^{-2}\beta(\kt{j}_i+K_{ij})+
\cosh^{-2}\beta(\kt{j}_i-K_{ij})\right) \nonumber \\
&&P_{ij}'=\frac{1}{2}\left(\cosh^{-2}\beta(\kt{j}_i+K_{ij})-
\cosh^{-2}\beta(\kt{j}_i-K_{ij})\right)
\; .
\eee
Due to (\ref{3.6}) two independent system of equations is obtained
from (\ref{3.5}). The static part:
\bee
 & m^{(1)}_i=\tanh\beta\kappat_i & \\
 & m^{(1)}_i=L_{ij}+P_{ij}m^{(1)}_j;\;
m^{(1)}_j=L_{ji}+P_{ji}m^{(1)}_i &\nonumber
\eee
can be reduced to the form
\be
m^{(1)}_i=\tanh\beta\kappat_i=\frac{L_{ij}+P_{ij}L_{ji}}{1-P_{ij}P_{ji}} \; .
\label{3.8}
\ee
The same equation is also obtained on the basis of cluster expansion of the
free energy~\cite{Yukhn} and can be written in the form
\be
m^{(1)}_i=\ang{S_i}_{\rho_i}=\ang{S_i}_{\rho_{ij}}
\ee
where the one-site $\rho_i$ and intracluster $\rho_{ij}$ density matrices are
defined as follows
\bee
&& \rho_i=\frac{\exp(-\beta H_i)}{Sp_{S_i}\exp(-\beta H_i)};\;
H_i=\kappat_iS_i \nonumber\\
&& \rho_{ij}=\frac{\exp(-\beta H_{ij})}{Sp_{S_iS_j}\exp(-\beta H_{ij})};\;
H_{ij}=\kt{j}_iS_i+\kt{i}_jS_j+K_{ij}S_iS_j
\eee
Multiplying (\ref{3.8}) by $X\ia$ and averaging obtained relation over
the sort configurations we get the equation for an average value of sort 
$\alpha$ spins $m^{(1)}\ia\equiv \avX{m^{(1)}_i X\ia}$
\bee
&& m^{(1)}\ia=c_\alpha\tanh\kappat\ia \nonumber\\
&& =\sum_\beta w\ab \frac{\sinh\beta(\kt{j}\ia+\kt{i}_{j\beta})+
		a\ab\sinh\beta(\kt{j}\ia-\kt{i}_{j\beta})}
{\cosh\beta(\kt{j}\ia+\kt{i}_{j\beta})+
 a\ab\cosh\beta(\kt{j}\ia-\kt{i}_{j\beta})}
\label{mia}
\eee
On assuming uniform field $\kappa_{i\alpha}\rightarrow\kappa_\alpha$
(UF: $\fr_{i\alpha}\rightarrow\bar{\varphi}_\alpha$,
$\kt{j}\ia\rightarrow\kappa_\alpha+(z-1)\bar{\varphi}_\alpha
\equiv\kappat_\alpha'$,
$\kappat\ia\rightarrow\kappa_\alpha+z\bar{\varphi}_\alpha
\equiv\kappat_\alpha$, $z$ is a first coordination number)
we obtain the known in corresponding static theories
(TCA~\cite{cmp1}, first order of cluster variation method~\cite{Sato})
relation: (\ref{mia}) becomes the set of $2\Omega$ equations 
for the same number of quantities -- $\Omega$ order parameters 
$m^{(1)}_\alpha$ and $\Omega$
effective fields $\bar{\varphi}_\alpha$ .

The dynamic linearized part of (\ref{3.5}) (here nonlinear on 
$\delta\kappa_{i,t}$ terms $L_{ij}'\; \delta m^{(1)}_{i,t}\; 
\delta\kt{i}_{j,t}$, $L_{ji}'\;\delta m^{(1)}_{j,t}\;\delta\kt{j}_{i,t}$ 
should be neglected) after time Fourier transformation and averaging over 
the sort configurations yields the Or\-n\-ste\-in-Zernike-type equation for 
the dynamic structure factor
$m^{(2)}\iajb(\omega)=
\frac{\delta m^{(1)}\ia(\omega)}{\delta \kappa\jb(\omega)}$
of the model. Spatial Fourier transformation presents its solution in the 
form
\bee
\left(\hat{m}^{(2)}(\vec{q},\omega)\right)^{-1}&=&
(1-z)\ptsed{\hafit{}}^{-1}+
   z\ptsed{\hafitz{}+\hafioo{}}^{-1} \nonumber\\
&+& \ptsed{\Piq{0}-\Piq{q}}
   \ptsed{\hafitz{}\ptsed{\hafioo{}}^{-1}\hafitz{} - \hafioo{}}^{-1}
        \label{StrFac}
\eee
where the matrix elements are
\bee
 &  & \ptsed{\hafit{}}\ab\stackrel{UF}{=}
\delta\ab c_\alpha\frac{1-\tanh^2\beta\kappat_\alpha}{D_\alpha} \nonumber\\
 &  & \ptsed{\hafitz{}}\ab\stackrel{UF}{=}
\delta\ab\sum_\gamma w_{\alpha\gamma}\frac{R_{\alpha\gamma} D_\gamma}{D_\alpha D_\gamma-P_{\alpha\gamma} P_{\gamma\alpha}} \nonumber\\
 &  & \ptsed{\hafioo{}}\ab\stackrel{UF}{=}
w\ab\frac{R_{\beta\alpha}P\ab}{D_\alpha D_\beta-P\ab P_{\beta\alpha}}\nonumber\\
 &  & \ptsed{\hat{m}^{(2)}(\vec{q},\omega)}\ab=
 \sum_j e^{i\vec{q}(\vec{R}_j-\vec{R}_i)} m^{(2)}\iajb(\omega)
\label{matr_el}
\eee
and
\bee
P\ab&=&\frac{1}{2}\left(\tanh\beta(\kappat_\alpha'+K\ab)-
\tanh\beta(\kappat_\alpha'-K\ab)\right) \\
R\ab&=&2a\ab \left[ 2a\ab(1+\cosh 2\beta\kappat_\alpha'\cosh 
2\beta\kappat_\beta')+\right. \nonumber\\
&&\left.(1+a\ab^2)(\cosh 2\beta\kappat_\alpha'+\cosh 
2\beta\kappat_\beta')\right] \times
\nonumber\\
&&\left[(1+a\ab^2+2a\ab\cosh 2\beta\kappat_\alpha'\right]^{-1} \times
\nonumber\\
&&\left[\cosh\beta(\kappat_\alpha'+\kappat_\beta')+
a\ab\cosh\beta(\kappat_\alpha'-\kappat_\beta')\right]^{-2}
 \nonumber
\eee
Factor $\Piq{q}$ has the form
\be
\Piq{q}=\sum_{j\in\pi_i}e^{i\vec{q}(\vec{R}_j-\vec{R}_i)};\;
\Piq{0}=z
\ee

\sect{Annealed model}
In this case protons (deuterons) can go from one site (hydrogen bond) to
another. Therefore the sort configuration changes in time. This motion
should be described by Kawasaki master equation \cite{Kawasaki}. But we
assume that the sort configuration changes slowly and for the frequencies of 
magnetic (or ferroelectric) dispersion quasispins see sort configuration at 
rest. Therefore the formula (\ref{StrFac}) remains valid. If the 
temperature changes quite slowly, $w\ab$ is a temperature-dependent quantity 
which can be found by the static theory \cite{cmp1}.

\sect{Discussion. Numerical results.}

The expression (\ref{StrFac}) is our target formula. Let us consider its
partial cases. All expressions will be given for UF case.

1) {\em Noninteracting system} ($K\ab=0\;\forall\alpha,\beta$) is a trivial
case.
\bee
&& m^{(1)}_\alpha=c_\alpha\tanh\beta\kappa_\alpha \\
&& m\ab^{(2)}(\vec{q},\omega)=\delta\ab
	\frac{c_\alpha (1-\tanh^2\beta\kappa_\alpha)}{1+i\omega\tau_\alpha^0}
\label{nonint}
\eee
All the spins are independent, so the structure factor are independent of
the wavelength. The relaxation times are equal to $\tau_\alpha^0$.

   2) {\em One-sort (ideal) system}. This case can be obtained in the limit
$c_1=1$. The single order parameter $m_1^{(1)}$ and the
effective field $\varphi_1$ satisfy the following equation
\be
m^{(1)}_1/c_1\equiv \sigma^{(1)}_1=\tanh\beta\kappat_1=
\frac{\sinh 2\beta\kappat_1'}{\cosh 2\beta\kappat_1'+a_{11}}
\label{(5.16a)}
\ee
where
\be
\kappat_1=\kappa_1+z\bar{\varphi}_1;\;\kappat'_1=\kappa_1+z'\bar{\varphi}_1;\;
z'=z-1;\;a=e^{-2K}
\ee
The dynamic structure factor 
$m^{(2)}_{11}(\vec{q},\omega)$
takes the form
\be
m^{(2)}_{11}(\vec{q},\omega)/c_1\equiv \sigma^{(2)}_{11}(\vec{q},\omega)=
\frac{\sigma^{(2)}_{11}(\vec{q},0)}{1+i\omega\tau(\vec{q})}
\label{DynStrFac} \; ,
\ee
\unitlength=1mm
\centerline{
\begin{picture}(120,142)
\put(0,142){\special{em:graph pqdep.pcx}}
\end{picture}	}
Figure 1: 
Pure system on plane square lattice ($z=4$). Static pair CF
$m_{11}^{(2)}(\vec{q},0)$, inverse relaxation time $1/\tau(\vec{q})$, real
and imaginary parts of dynamic pair CF $m_{11}^{(2)}(\vec{q},\omega)$ at
different wavevectors: $\Piq{q}$ equals $4$ ($\vec{q}=0$), $2$, $0$, $-2$,
$-4$ (the last corresponds to the vertices of the first Brillouin zone).
\\[1.2em]
where
\be
\sigma^{(2)}_{11}(\vec{q},0)=
\ptsed{1-\ptsed{\sigma^{(1)}_1}^2}\frac{1-P^2}{1+z'P^2-P\Piq{q}}
\label{StatStrFac}
\ee
\[
P=\frac{1-a^2}{1+a^2+2a\cosh 2\beta\kappat'}
\frac{1}{1-\ptsed{\sigma^{(1)}_1}^2} \; ,
\]
and $\sigma_{11}^{(2)}(\vec{q},0)$ coincides with a spatial 
Fourier-transform of
the static correlation function $\ang{S_iS_j}^c_H$, which was founded with
cluster expansion of the free energy within TCA in the ref.~\cite{Yukhn}.
Result (\ref{StatStrFac}) is accurate in the one-dimensional case.
Relaxation time
\be
\tau(\vec{q})=\tau_1^0\left(1-\Piq{q}
\frac{P}{1+z'P^2}\right)^{-1}
\stackrel{T>T_c}{\longrightarrow}\tau^0\ptsed{1-\Piq{q}\frac{\tanh\beta
K}{1+z'\tanh^2\beta K} }^{-1}
\label{InvRelTime}
\ee
is a decreasing function of $\vec{q}$. Note, that known exact result
\cite{SandK} (zero external field, ideal one-dimensional system) is a partial
case of the formula (\ref{DynStrFac}).

Figure 1 
demonstrates the temperature and $\vec{q}$-dependence of
the static structure factor $m_{11}^{(2)}(\vec{q},0)$ and the relaxation time
$\tau(\vec{q})$ of the system on the plane square lattice ($z=4$).
Five lines represent five wavevectors: $\Piq{q}=4$, $2$, $0$, $-2$, $-4$.
In this figure (as well as in the formula (\ref{InvRelTime})) one can see an 
interesting symmetry of inverse relaxation time: $tau_1^0/tau(\vec{q})$ 
line at $\Piq{q}$ is mirror reflection of the line at $-\Piq{q}$.
The figure 1 
shows also a behaviour of real and imaginary parts 
of the dynamic structure factor (at the frequency $\omega=0.1/\tau^0_1$).

3. {\em Diluted system} (interacting spins (sort 1) are diluted by the
noninteracting (nonmagnetic, nonferroelectric) impurities:
$K_{1\alpha}=0$ for all $\alpha\ne 1$) is
interesting due to percolation phenomena in such system: below some
concentration of sort 1 spins there is no infinite cluster of interacting
spins and ordered phase does not appear. The mean field
approximation (MFA) fails to describe percolation phenomena \cite{Sato}.
At zero external field within MFA $\sigma^{(1)}_1$ and pair CF per 
interacting spin $\sigma^{(2)}_{11}$ depend on temperature, concentration and
interaction strength only via $T/T^{MFA}_c$ ($T^{MFA}_c=zc_1K_{11}/k_b$):
$\sigma^{(1)}_1=\sigma^{(1)}_1(T/T^{MFA}_c)$,
$\sigma^{(2)}_{11}=\sigma^{(2)}_{11}(\vec{q},\omega,T/T^{MFA}_c)$.
Therefore MFA predicts that $\sigma^{(1)}_1$
reaches its saturation value ($\sigma^{(1)}_1=1$ at $T=0$) at any
concentration of interacting spins, and $m^{(2)}_{11}(\vec{q},\omega)$
shows a single relaxation time behavior.
In TCA $T_c={2K_{11}\over k_B\ln\frac{z'w_{11}+c_1}{z'w_{11}-c_1}}$,
\be
m^{(1)}_1=c_1\tanh\beta\kappat_1=
w_{11}\frac{\sinh 2\beta\kappat_1'}{\cosh 2\beta\kappat_1'+a_{11}}+
(c_1-w_{11})\tanh\beta\kappat_1'
\label{(5.18a)}
\ee
and this takes place for the case $w_{1\alpha}=0$ only, otherwise at zero
temperature $\sigma^{(1)}_1<1$ (due to the presence of finite size clusters
of interacting spins) and becomes zero if $c_1\ge z'w_{11}$ (below
percolation point). In the case of complete chaos ($w\ab=c_\alpha c_\beta$)
TCA predicts percolation concentration $c_p=(z-1)^{-1}$. We note that the
Effective Field Approximation of Kaneyoshi et.al. \cite{Kaneyoshi}
(magnetization and static susceptibility) also reproduces this property
of the diluted system.

Dynamic pair CF contains three terms
\be
 m^{(2)}_{11}(\vec{q},\omega)= \frac{A_0}{1+i\omega\tau_1^0}+
 \frac{A_+(\vec{q})}{1+i\omega\tau_+(\vec{q})}+
 \frac{A_-(\vec{q})}{1+i\omega\tau_-(\vec{q})}
\label{4.10}
\ee
where
\bee
&&A_0=\frac{c_1 F_1^{(2)}w_{12}R_{12}}{zc_1F_1^{(2)}-z'w_{12}R_{12}}
\stackrel{T>T_c}{\longrightarrow}\frac{c_1w_{12}}{c_1+z'w_{11}}  \nonumber\\
&&A_\pm(\Piq{q})=\frac{B_\pm(\Piq{q})}{1-D_\pm(\Piq{q})};\;
\tau_\pm(\vec{q})=\frac{\tau_1^0}{1-D_\pm(\Piq{q})} 
\eee
\newpage
\unitlength=1mm
\centerline{
\begin{picture}(120,185)
\put(0,185){\special{em:graph qdep.pcx}}
\end{picture}}
\noindent Figure 2: 
Diluted system ($c_1=0.6$, $w_{11}=c_1^2$) on plane square lattice
(percolation concetration $c_p=1/3$). Amplitudes $A_+$, $A_-$, $A_0$,
inverse relaxation times $1/\tau_+(\vec{q})$, $1/\tau_-(\vec{q})$, and
dynamic pair CF $m_{11}^{(2)}(\vec{q},\omega)$ at
different $\vec{q}$: $\Piq{q}=4, 2, 0, -2, -4$. 
\newpage
\unitlength=1mm
\centerline{
\begin{picture}(120,140)
\put(0,140){\special{em:graph chiw2.pcx}}
\end{picture}}
\noindent Figure 3: 
Diluted system. Temperature and frequency dependence of dynamic
pair CF $m_{11}^{(2)}(\vec{q},\omega)$ at $\vec{q}=0$. \\[1.2em]
\[
B_\pm(\Piq{q})=\pm\frac{c_1 F_1^{(2)}}{PD_\pm(\Piq{q})
\sqrt{{\cal D}(\Piq{q})}}
\ptsed{\ang{R_1}^2 D_\pm^2(\Piq{q})-(Pw_{12}R_{12})^2}
\]
and
\bee
\ang{R_1}&=&w_{11}R_{11}+w_{12}R_{12} \nonumber\\
D_\pm(\Piq{q})&=&P
\frac{c_1F_1^{(2)}w_{11}R_{11}\pm\sqrt{{\cal D}(\Piq{q})}}
{2\ang{R_1}\ptsed{zc_1F_1^{(2)}-z'\ang{R_1}}} \nonumber\\
{\cal D}(\Piq{q})&=&\ptsed{c_1F_1^{(2)}w_{11}R_{11}\Piq{q}}^2 \nonumber\\
&+&4\ang{R_1}\ptsed{zc_1F_1^{(2)}-z'\ang{R_1}}w_{12}R_{12}
\ptsed{zc_1F_1^{(2)}-z'R_{12}} 
\eee
\newpage
\unitlength=1mm
\centerline{
\begin{picture}(120,187)
\put(0,188){\special{em:graph ces_a.pcx}}
\end{picture} }
\noindent Figure 4: 
Spontaneous polarization $P$, inverse permittivity $1/\varepsilon$ and 
inverse relaxation time $1/\tau$ of the ferroelectric $Cs(H_{1-x}D_x)_2PO_4$ 
for different degrees of deuteration.
Squares correspond to experimental data of \cite{Deguchi1}, bold lines on 
the lower picture connect experimental points from the ref. \cite{Deguchi2}. 
Thick lines -- theoretical results for annealed system with parameters given 
in table 1 and concentrations $x=0$, 0.18, 0.54, 0.88, 1. 
\newpage
\unitlength=1mm
\centerline{
\begin{picture}(120,190)
\put(0,190){\special{em:graph ces_q.pcx}}
\end{picture} }
\noindent Figure 5: 
Comparison of experiments and quenched model with assumption $w\ab=c_\alpha 
c_\beta$ (complete chaos) (see caption to the previous figure).
Thick lines -- theoretical results for quenched system with parameters given 
in table 1 and concentrations $x=0$, 0.12, 0.47, 0.83, 1.  
\newpage
\bee
F_1^{(2)}&=&c_1(1-(\sigma^{(1)}_1)^2) \nonumber\\
R_{12}&=&1-\tanh^2\beta\kappat_1' \nonumber\\
R_{11}&=& 4a\frac{1+a\cosh 2\beta\kappat_1'}
{(\cosh 2\beta\kappat_1'+a)(1+a^2+2a\cosh 2\beta\kappat_1')} \nonumber\\
P&=&\frac{1}{2}\left[\tanh\beta(\kappat_1'+K)-\tanh\beta(\kappat_1'-K)\right]
\eee
The first term of (\ref{4.10}) has the (\ref{nonint})-like behavior
(relaxation time equals $\tau_1^0$), so we believe it to be a contribution
of isolated spins (i.e. the spins, which all neighbours are noninteracting).
The last two terms of (\ref{4.10}) have the following symmetry
\bee
&&B_\pm(\Piq{q})=B_\mp(-\Piq{q});\;A_+(-z)=A_-(z)=0 \nonumber\\
&&D_\pm(\Piq{q})=-D_\mp(-\Piq{q});\;D_+(\Piq{q})>0;
\;D_-(\Piq{q})<0
\eee
\begin{table*}
\centerline{
\begin{tabular}{|l|cc|}\hline
      		  &  quenched & annealed\\ \hline
$K_{11}/k_B$, $K$ &     \multicolumn{2}{c}{610}\vline\\
$K_{12}/k_B$, $K$ &     500   & 500\\
$K_{22}/k_B$, $K$ &     \multicolumn{2}{c}{390}\vline\\
$J_{11}/k_B$, $K$ &     \multicolumn{2}{c}{2.85}\vline\\
$J_{12}/k_B$, $K$ &     3.2   & 1.6\\
$J_{22}/k_B$, $K$ &     \multicolumn{2}{c}{1.06}\vline\\
$(V_{11}+V_{22}-2V_{12})/k_B$, $K$      &     --    & -1000\\
$I\ab/k_B$, $K$   &     --    & 0\\
$\mu_1$, $e.s.u.\times cm$& \multicolumn{2}{c}{2.18$\times 10^{-18}$}\vline\\
$\mu_2$, $e.s.u.\times cm$& \multicolumn{2}{c}{1.87$\times 10^{-18}$}\vline\\
$\tau^0_1$, $s$ & \multicolumn{2}{c}{3$\times 10^{-14}$}\vline\\
$\tau^0_2$, $s$ & \multicolumn{2}{c}{5.5$\times 10^{-15}$}\vline\\ \hline
\end{tabular}
}
\caption{
The model parameters describing experimental data on 
the pseudo-one-dimensional ferroelectric $Cs(H_{1-x}D_x)_2PO_4$:
short-range coupling $K\ab$; long-range interaction $J\ab$;
effective dipole electric moments $\mu_\alpha$ associated with proton or 
deuteron on hydrogen bond; parameters of master equation --
relaxation times $\tau^0_\alpha$ of noninteracting protons or deuterons.
$V\ab$, $I\ab$ are parameters describing nonexchange interactions between 
pseudospins [2]. 
Deuterons correspond to the pseudospins of sort 1.
}
\end{table*}
The relaxation time $\tau_-$ is always
less then $\tau_1^0$ while $\tau_+$ is always greater then $\tau_1^0$ and
slows down at the critical point ($\tau_+(\vec{0})
\stackrel{T\rightarrow T_c}{\longrightarrow}\infty$, $A_+(\Piq{0})
\stackrel{T\rightarrow T_c}{\longrightarrow}\infty$).
Both $\tau_-$ and $\tau_+$ increase with the approaching to the center of the
first Brillouin zone ($\vec{q}\rightarrow 0$). The amplitude $A_-$ is
equal to $0$ in the centre of the first Brillouin zone ($\Piq{q}=-z$)
while $A_+$ vanishes at
$\Piq{q}=-z$ (in the vertices of the first Brillouin zone).

In the case $w_{11}=c_1$ ($w_{1\alpha}=0$ $\forall\alpha\ne 1$,
all interacting spins belong to one
infinite cluster), $A_0$ is equal to zero, $A_-$ equals zero at
$\pi(\vec{q})>0$ and $A_+$ is equal to zero at $\pi(\vec{q})<0$;
$\sigma^{(1)}_1$ and pair CF are given by the formulae (\ref{(5.16a)}),
(\ref{DynStrFac}).

Figure 2 shows the pair CF (\ref{4.10}) of the diluted system on
the plane square lattice ($z=4$). In the figure 3 one can observe
that presence of many relaxation times in dynamic pair CF of the diluted
system can obviously be seen at small temperatures and frequencies.
Abscissas of maxima of $-$Im $m_{11}^{(2)}(\vec{0},\omega)$ approximately
equal inverse relaxation times:
$\omega_0\approx 1/\tau^0_1$, $\omega_+\approx 1/\tau_+(\vec{0})$.
Specific temperature dependence of $m_{11}^{(2)}(\vec{0},\omega)$ is
connected with nonmonotonouos behavior of $\tau_+(\vec{0})$.

\sect{Experiments}
      One can see that TCA reflects essential
properties of the diluted system. It seems to us
that quality of the approximation is even higher for the system without
noninteracting impurities ($K\ab\ne 0$) because such "effective field"
theories are more appropriate in the case of small
fluctuations in a system considered.
Obtained within TCA characteristics can be used
in order to describe a variety of experimental data
on partially deuterrated ferroelectrics with hydrogen bonds
and alloys of magnets. For example, good fit of TCA results for the ideal
(one-sort) system to experimental data on the ferroelectric dispersion in
pseudo-one-dimensional ($z=2$) ferroelectrics $CsH_2PO_4$ and its deuterated
analog $CsD_2PO_4$ has been reached in the ref. 
\cite{Levitskii_disp,Levitskii_disp_r}. The
many-sort theory developed here is applicable to partially deuterated
compound -- $Cs(H_{1-x}D_x)_2PO_4$. We obtain a good agreement of TCA
results on annealed model and results of dielectric measurements 
\cite{Deguchi1,Deguchi2} (figure 4). Fit of the quenched model (figure 5) to 
experimental data is not so good. The model parameters are shown in the 
table 1. More information about an actual type of the system can be given by 
measurement in the low-temperature region, where significant differences 
between behaviours of the annealed and quenched models exist, and by the 
wavevector-dependent experiments (e.g., neutron scattering).

\section*{Appendix}
\setcounter{equation}{0}\renewcommand{\theequation}{A\arabic{equation}}
\begin{appendix}

Here we consider a system with the Hamiltonian
\be
 -{\cal H}/k_BT = \frac{1}{2}\sum\iajb K\iajb S\ia S\jb+
\frac{1}{2}\sum\iajb J\iajb S\ia S\jb+\sum\ia\Gamma\ia S\ia \; ,
\ee
where $K\iajb$, $J\iajb$ are pair interaction parameters, $\Gamma\ia$ external
field acting on the spin of sort $\alpha$ on the site $i$. The interaction
$K\iajb$ is same as in (\ref{Hamilt})-(2.3), but $J\iajb$ is assumed to be of
long-range type, i.e. not restricted to the nearest neighbour sites.
The Hamiltonian (A1) can be written as
\bee
&& {\cal H}=\,^{MFA}H-\frac{k_BT}{2}\sum\iajb J\iajb \Delta^Lm^{(1)}\ia \Delta^Lm^{(1)}\jb \label{(A2)}\\
&& ^{MFA}H=H+\frac{k_BT}{2}\sum\iajb J\iajb\, ^Lm^{(1)}\ia\, ^Lm^{(1)}\jb \nonumber\\
&&\Delta^Lm^{(1)}\ia=S\ia-\,^Lm^{(1)}\ia \; , \nonumber
\eee
where $H$ is the Hamiltonian (\ref{Hamilt}) in which the field $\kappa$ is
taken in the form
\bee
&&\kappa\ia=\Gamma\ia+\sum\jb J\iajb\, ^Lm^{(1)}\jb \; .
\eee
MFA assumes
\be
^Lm^{(1)}\ia=\avX{\ang{S\ia}_{{\cal H},t}}
\ee
and ignores last term of (\ref{(A2)}):
\be
{\cal H}\longrightarrow\,^{MFA}H=H+\frac{k_BT}{2}\sum\iajb J\iajb \,^Lm^{(1)}\ia\,^Lm^{(1)}\jb \; .
\label{(A5)}
\ee
One can see that formalism of the section~3 is applicable to the system with
the Hamiltonian (\ref{(A5)}). The equation of motion has the same form
\be
D_i\ang{S_i}_{{\cal H},t}=\ang{\tanh\beta\varepsilon_i}_{{\cal H},t};
\qquad \varepsilon_i=\sum_jK_{ij}S_j+\kappa_i \; ,
\ee
but now the field $\kappa_i$ includes molecular field (see (A3)) and thus
couples the selfconsistency parameter $^Lm^{(1)}\ia$. Then one can find that
\be
^Lm^{(1)}\ia=m^{(1)}\ia \; ,
\ee
where $m^{(1)}\ia$ is defined by the equation (\ref{mia}) and
\bee
^Lm\iajb^{(2)}(\omega)& \equiv &\fdif{^Lm^{(1)}_{i\alpha}(\omega)}{\Gamma_{j\beta}(\omega)}
=\sum_{l\gamma}\fdif{m^{(1)}_{i\alpha}(\omega)}{\kappa_{l\gamma}(\omega)}
\fdif{\kappa_{l\gamma}(\omega)}{\Gamma_{j\beta}(\omega)} \nonumber\\
	&=&\sum_{l\gamma}m_{i\alpha,l\gamma}^{(2)}(\omega)
\ptsed{\delta_{lj}\delta_{\gamma\beta}+\sum_{k\delta}J_{l\gamma,k\delta}
\,^Lm^{(2)}_{k\delta,j\beta}(\omega)} \; .
\label{(A8)}
\eee
(\ref{(A8)}) leads to the following expression for the structure factor
\be
\ptsed{^L\hat{m}^{(2)}(\vec{q},\omega)}^{-1}=
\ptsed{\hat{m}^{(2)}(\vec{q},\omega)}^{-1}-\hat{J}(\vec{q}) \; ,
\ee
where $\hat{m}^{(2)}(\vec{q},\omega)$ is defined by (\ref{StrFac}) and
$\ptsed{\hat{J}(\vec{q})}\ab$ is a Fourier-transform of $J\iajb$.
\end{appendix}

\end{document}